\documentclass[conference]{IEEEtran}

\usepackage[]{algorithm2e}
\usepackage{graphicx}
\usepackage{amsmath}

\begin{document}

\title{Stay-point Identification as Curve Extrema}

\author{\IEEEauthorblockN{Georgios Stylianou}
\IEEEauthorblockA{Department of Computer Science\\
European University Cyprus\\
Engomi, Cyprus\\
Email: g.stylianou@euc.ac.cy}
}
\IEEEpeerreviewmaketitle
\maketitle

\begin{abstract}
In a nutshell, stay-points are locations that a person has stopped for some amount of time. Previous work depends mainly on stay-point identification methods using experimentally fine tuned threshold values. These behave well on their experimental datasets but may exhibit reduced performance on other datasets.  

In this work, we demonstrate the potential of a geometry-based method for stay-point extraction. This is accomplished by transforming the user's trajectory path to a two-dimensional discrete time series curve that in turn transforms the stay-points to the local minima of the first derivative of this curve. 

To demonstrate the soundness of the proposed method, we evaluated it on raw, noisy trajectory data acquired over the period of $28$ different days using four different techniques. The results demonstrate, among others, that given a good trajectory tracking technique, we can identify correctly $86\%$ to $98\%$ of the stay-points.
\end{abstract}

\section{Introduction}
The knowledge of a person's spatial whereabouts is an important component of a ubiquitous system. In the past years, several researchers have proposed various methods for automatic collection of a user's spatial trajectories using smartphone applications~\cite{Chon14,Kjaergaard11,Stylianou15} running in the background. Given appropriate algorithms for processing this spatial data in order to extract useful information such as stay-points and user patterns, important applications emerge. Examples of such applications include predicting a user's movement~\cite{Herder14}, tourism~\cite{Horanont14}, extracting individual life patterns~\cite{Ye09} and  recommending locations and activities~\cite{Zheng10a}.

In this work, we are interested in the user's stay-points. A \textit{stay-point} is a location where the user has stopped for some amount of time. This could be the user's home or work where the duration is large or could be the user's children's school where a daily transportation is done. Even though the stay-point identification problem has been previously addressed by several researchers, the problem was mainly tackled as a point clustering problem where usually existing clustering techniques were used. An important aspect of previous work, was the identification of the maximum distance between two points such that they are considered to be in the same location and the minimum time between them such that they are labeled as stay-points~\cite{Ashbrook03, Kang05, Palma08, Zheng10, Horanont14, Ranvier15}. In contrast, recent work~\cite{Zhou14, Yadav14, Thomason15} attempts to reduce the dependency on these thresholds but introduces other thresholds.  Another problem is that confusion is generated to the reader as there is no agreement on the values of distance and time thresholds because their values vary significantly among different works. In addition, despite the fact that the exhibited performance of previous work was quite good, demonstrating stay-point identification rates of over $90\%$, this performance is mainly achieved after experimental fine tuning of the different threshold values such that they fit very well their datasets.

\begin{figure}[t]
  \centering\begin{tabular}{c}
     \begin{minipage}{6cm}
     \centering
     \includegraphics[width=6cm]{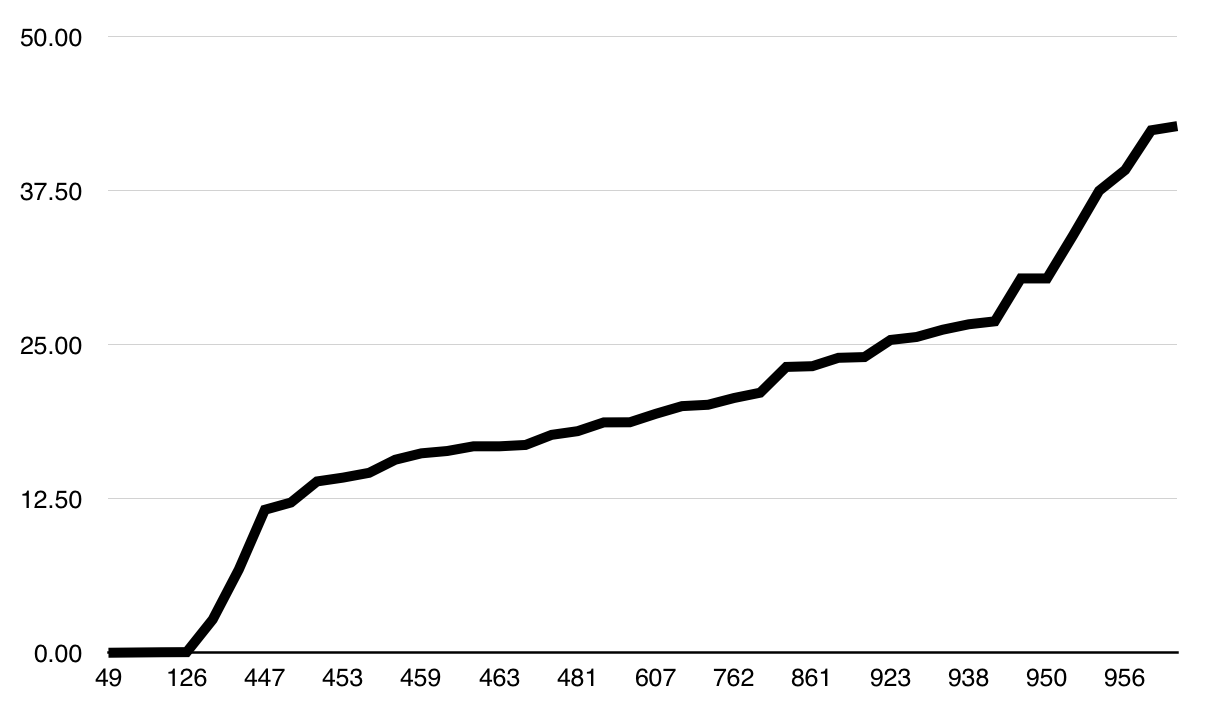}
    \end{minipage}
\\
     \begin{minipage}{6cm}
     \centering
     \includegraphics[width=6cm]{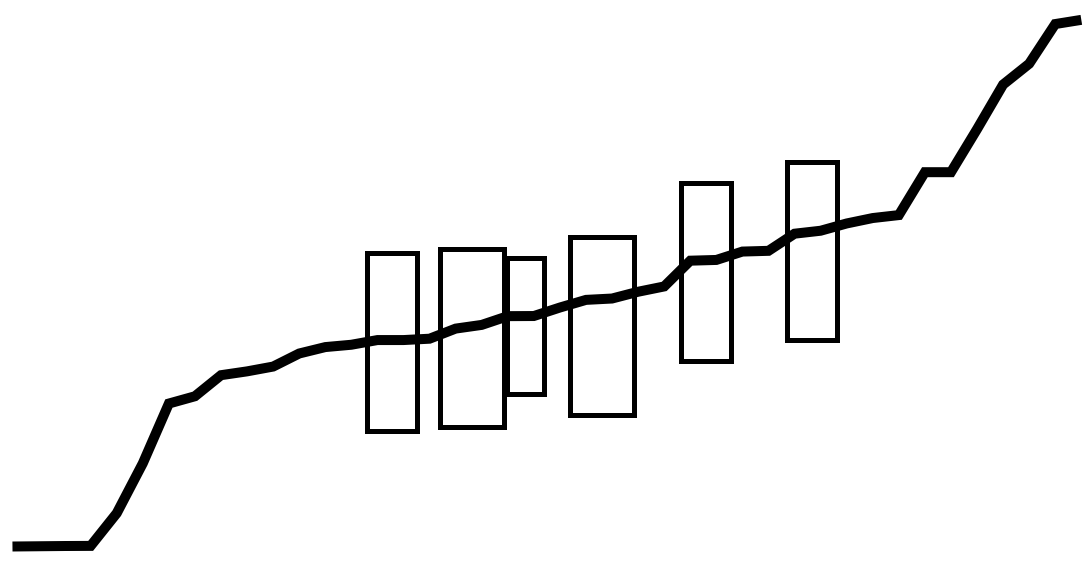}
    \end{minipage}

\end{tabular}
\caption{Method overview. We transform the user's trajectory for a single day to a 2D spatial curve. Using extrema extraction, we identify the user's stay-points; these are depicted using rectangles.}
\label{fig:methodOverview}
\end{figure}
It is evident that an appropriate method that reduces significantly the dependence on experimentally deduced threshold values or a method that requires little or no fine tuning does not exist. In contrast, we propose a different perspective towards stay-point identification as depicted in figure~\ref{fig:methodOverview}. Instead of directly processing the recorded raw spatial trajectories, we first transform each daily spatial trajectory to a two-dimensional ($2D$) discrete spatial curve. Now the problem of stay-point identification is transformed from a thresholding problem to an extrema extraction problem. Since, the spatial curve models displacement over time and it's first derivative models speed over time, one can intuitively see it as unifying distance and time thresholds into a single threshold. However this new threshold required to identify an extremum has a well-posed numerical value.
Among others, we show that the proposed method produces good results on both sparse and dense spatial trajectory data, does not depend in any way on the uniformity of the data points and it is stable over large positioning errors that raw trajectory data exhibit.

This article has two main contributions:
\begin{itemize}
\item We propose a novel method for stay-point identification. This method tackles the problem from a completely different perspective. In this same time it eliminates dependency on threshold values that are experimentally deduced and are good only on specific datasets. 
\item The proposed method is very stable over different trajectory densities, works well on uniformly and non-uniformly spaced raw data points.
\end{itemize}

\section{Related Work}
\label{sec:relatedwork}
Past work on stay-point extraction can be divided into two groups. The work that depends on time and distance thresholds~\cite{Ashbrook03,Hariharan04, Kang05, Palma08, Zheng09, Zheng10,  Nishida14, Horanont14, Ranvier15, Pavan15} and the work that attempts to eliminate this dependency~\cite{Zhou14, Yadav14, Thomason15}. 

One of the first methods was developed by Ashbrook and Starner~\cite{Ashbrook03} that proposed that stay-points are points that have stay time of at least $10$ minutes and used the k-means clustering algorithm in order to discover the optimal distance that reveals stay-points. The next year, Hariharan et al.~\cite{Hariharan04} used the same thresholds but they proposed that their values should be tuned according to the application. Nevertheless, they show results for distance and time threshold of $50$ meters and $10$ minutes, respectively. Kang et al.~\cite{Kang05} suggest that a good value for the distance threshold is $30-50m$ and for time is $6$ minutes for long stays and proposed a time threshold of $100-150$ seconds for quick stays. Even though Palma et al.~\cite{Palma08} still use a distance threshold they try to compute it by computing the quantile function - the inverse of the cumulative distribution function - of the gaussian curve for the arithmetic mean and standard deviation of the list of distances between consecutive points. The quantile function and a time threshold value, proposed to be set to two minutes, are used to improve the DBSCAN clustering algorithm for the purpose of stay-point identification. Later, Zheng et al.~\cite{Zheng10,Zheng09} proposed to identify stay-points using a time threshold of $20$ minutes and a distance threshold of $200$ meters. Also, Nishida et al.~\cite{Nishida14} propose a $20$ meter distance threshold and $5$ minute time threshold. Horahont et al.~\cite{Horanont14} consider the distance threshold of $196$ meters and the time threshold of $14$ minutes to be good as identified out of experiments with $15$ people. Finally, Pavan et al.~\cite{Pavan15} decide that for stay-point labeling the distance threshold is $50$ meters and the time threshold is $50$ seconds. They also introduce a speed threshold defined as the median of all speeds reached within the trajectory data.

In contrast, other researchers~\cite{Zhou14, Yadav14, Thomason15} proposed different approaches towards eliminating this dependency. Zhou et al.~\cite{Zhou14} in order to detect a stay-point introduce another threshold denoted as $\mathbf{K}$. This is used in a time stamp clustering algorithm to pre-process the trajectory data such that a classical clustering algorithm such as k-means and DBSCAN produce better results. In addition, the authors experiment with a time threshold. 

Yadav et al.~\cite{Yadav14} record Cell Ids and use a graph based clustering algorithm in order to identify stay points. The graph is built such that the cell ids are the vertices and two cell ids are incident if their time difference is less than $\alpha$; this ensures successive cell ids are connected. Then they cluster vertices, within the same cell id, using their edge weight and vertex degree which are controlled by two arbitrary parameters $\eta$ and $\eta'$, respectively. Each cluster within a Cell represents a stay-point. 

Thomason et al.~\cite{Thomason15} propose a gradient-based visit extractor algorithm. This algorithm works as follows: insert all points into a buffer until the user has moved more than some threshold (computed via the gradient) or the time difference between the first point and the last point in the buffer is greater than another time threshold; then if there is some duration between the first and last points in the buffer, then a stay-point is identified. The threshold for the gradient, that controls the distance the user moved, depends on two parameters, which are experimentally tuned per dataset.

It is obvious that there is no consensus on the number of thresholds and their the best values. In the previous work, the values for time threshold range from $50$ seconds to $20$ minutes for long stays and the values for distance threshold range from $20$ meters to $200$ meters. Even though these numbers came out of experimental results with groups of people, these results are confusing when studying, evaluating or even attempting to implement a previous method. In addition, we believe that the stay-point identification problem can be more effectively tackled as a curve extremum problem than a clustering problem for reasons including reduction of the two main thresholds of time and distance to a single more intuitive threshold and the simplification of implementation.
Furthermore, we show in our experimental results (see table \ref{table:inflectionPointDurations}) that these hard threshold values lead to false identification of stay-points.

\section{Proposed Method}
\label{sec:proposedMethod}
We propose to transform the problem from a spatial to a geometry problem such that the stay-points become curve extrema. In this section, we provide the problem definition, describe the problem transformation and explain how we identify the stay-points.

\subsection{The Problem}
\label{subsec:theProblem}

\textbf{Objective}. Given a spatial trajectory (of a person) for a single day, the objective is to identify all the stay-points of this person. 

\textbf{Input}. The spatial trajectory is represented as the set $S=\{loc_i=(t_i,\mathbf{l}_i)|\,i=1...n\}$, where  $\mathbf{l}_i=(latitude,longitude)$ are geographical coordinates and $t_i$ (timestamp) is the date/time that $\mathbf{l}_i$ is recorded. 

\textbf{Output}.  The stay-points, denoted as $\mathbf{s}_i$, which are the locations the user has stayed for some amount of time.

\subsection{Problem Transformation}
\label{subsec:problemTransformation}
As already mentioned, we tackle the problem of stay-point identification under a totally different perspective with respect to the previous work. The first step is the transformation of the spatial trajectory to a spatial curve as it is depicted in figure~\ref{fig:methodOverview}.

The \textit{spatial curve} is a $2D$ increasing trajectory curve $C$ composed of points  $\mathbf{p}_i=(x_i,y_i)$, where $x_i$ is the elapsed time since midnight (i.e. the beginning of the day) in minutes and $y_i$ is the total displacement, in kilometers, until minute $x_i$.

The transformation is achieved via a bijective map $M$ (eq.~\ref{eq:map}), which is composed of the functions $f$ (eq.~\ref{eq:f},~\ref{eq:f_explained}) and $g$ (eq.~\ref{eq:g},~\ref{eq:g_explained}). The function $f$ (eq.~\ref{eq:f_explained}) transforms the timestamp $t_i$ to the $x_i$-coordinate of a point and the function $g$ (eq.~\ref{eq:g_explained}) transforms the position $\mathbf{l}_i$ to the $y_i$-coordinate of a point. 

\begin{gather}
\label{eq:map}
M: (t_i,\mathbf{l}_i)\rightarrow (x_i,y_i) \\
\label{eq:f}
f:t_i\rightarrow x_i\\
\label{eq:g}
g:\mathbf{l}_i\rightarrow y_i \\
\label{eq:f_explained}
\left. f(t_i)=t_i-midnight\right\vert_{\mbox{in minutes}}\\
\label{eq:g_explained}
g(l_i)=\sum_{i=2}^{n} distance(l_i,l_{i-1})\mbox{, }g(l_1)=0 
\end{gather}

The function $g$ transforms the geographical coordinate $\mathbf{l}_i$ to the cumulative distance from the first geographical coordinate $\mathbf{l}_1$. 

\begin{figure}[h]
  \centering\begin{tabular}{c}
     \begin{minipage}{7cm}
     \centering
     \includegraphics[width=7cm]{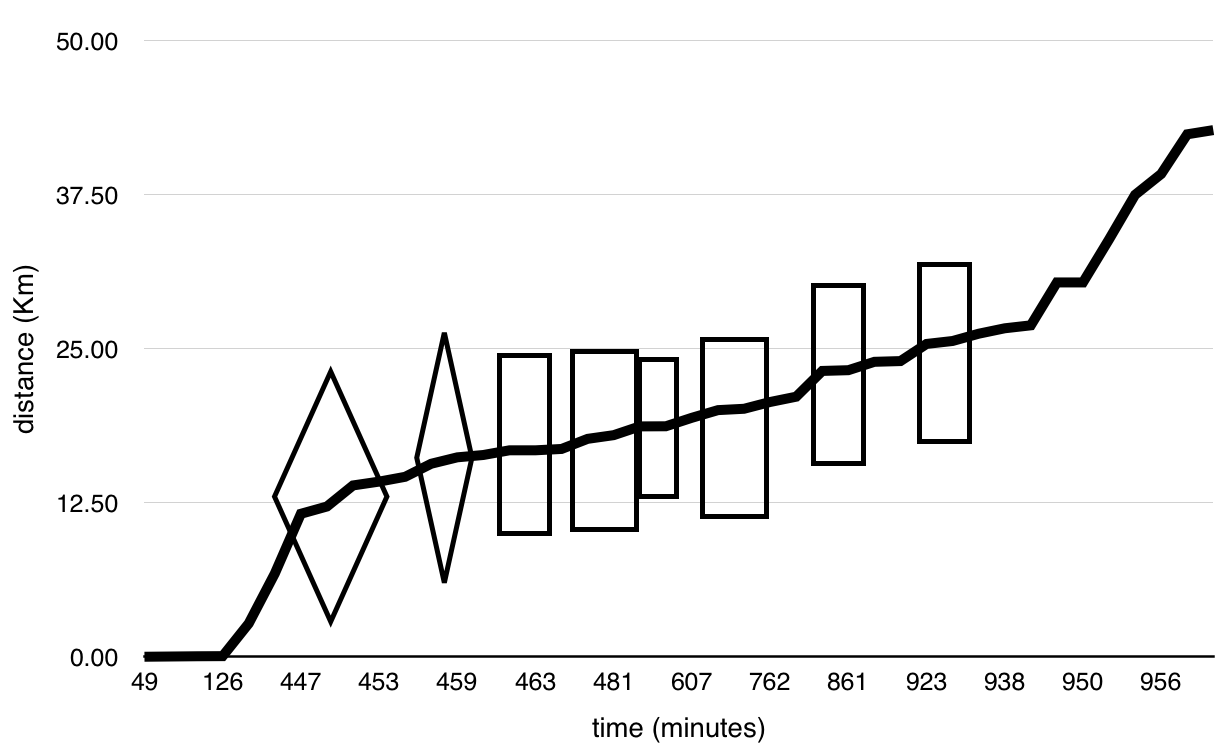}
    \end{minipage}
\\
(a)\\
     \begin{minipage}{7cm}
     \centering
     \includegraphics[width=7cm]{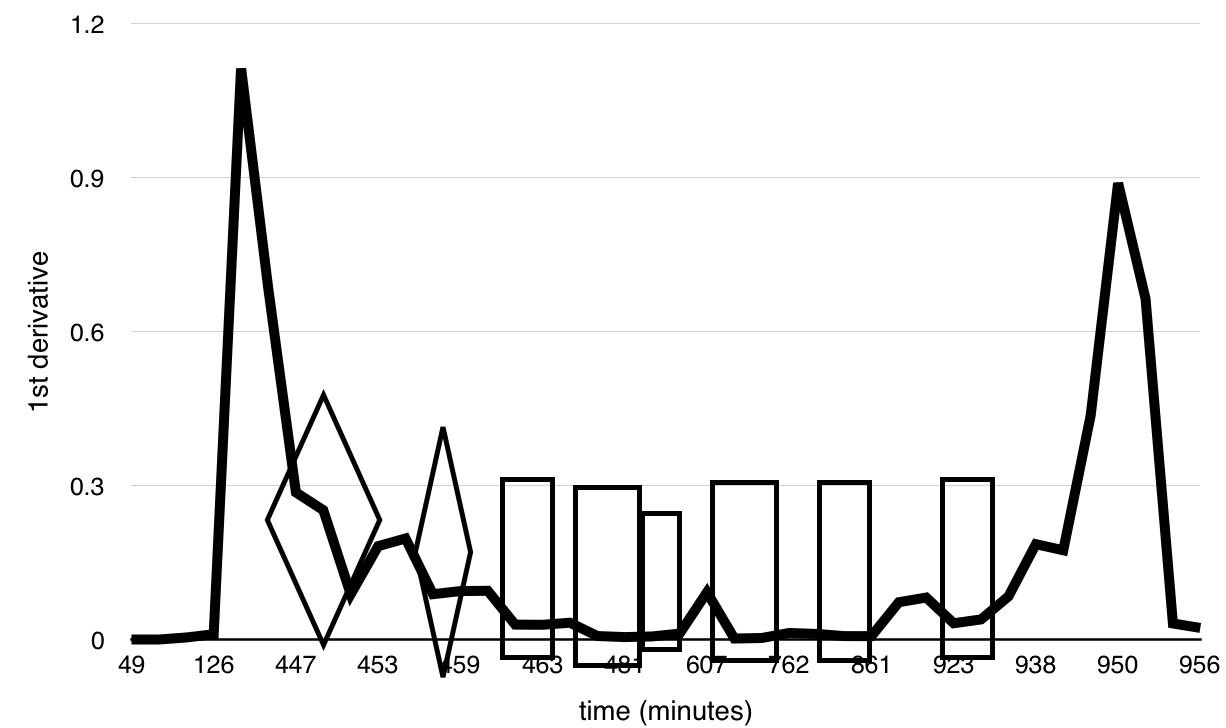}
   \end{minipage} 
  \\
 (b) \\
     \begin{minipage}{7cm}
     \centering
     \includegraphics[width=7cm]{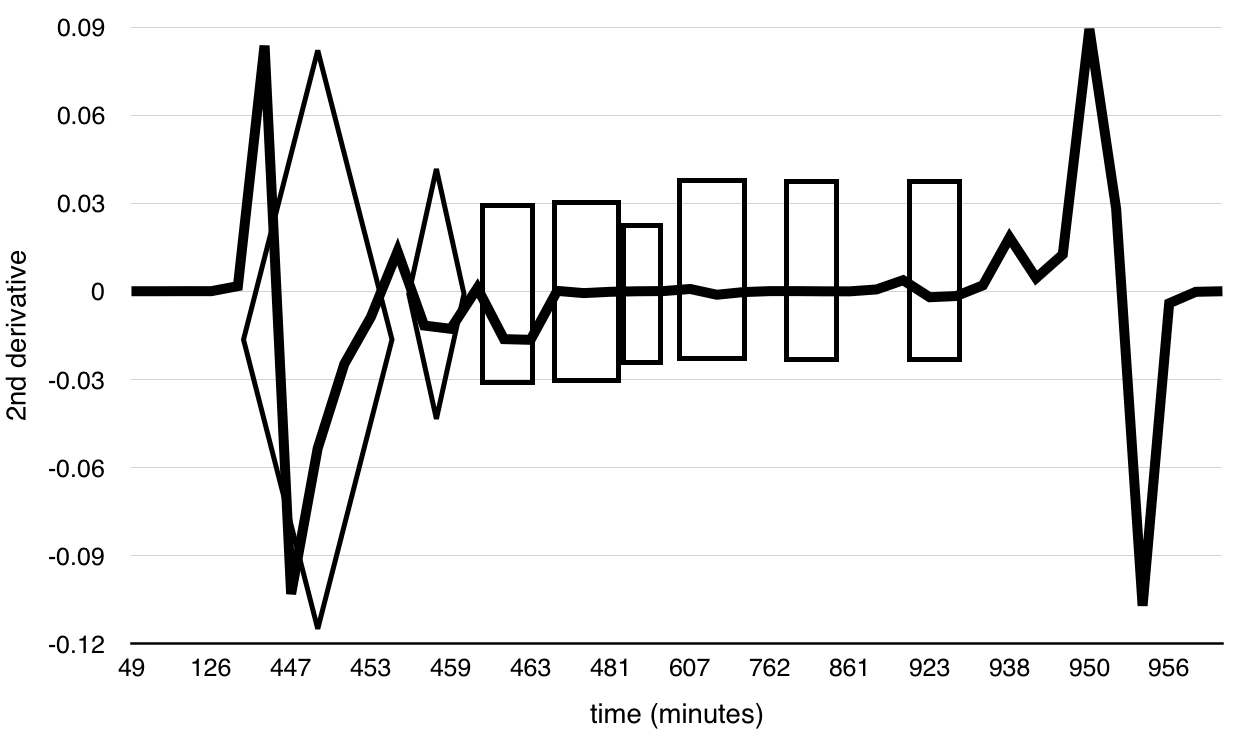}
    \end{minipage}
\\
(c) 
\end{tabular}
\caption{Spatial curve overlayed with inflection points (diamonds) and stay-points (rectangles). (a) spatial curve, (b) $1^{st}$ derivative spatial curve, (c) $2^{nd}$ derivative spatial curve.}
\label{fig:spatialCurveWithDerivatives}
\end{figure}
\subsection{Stay-point Identification}
\label{subsec:stay-pointIdentification}
Given the resulting $2D$ spatial curve, we are in place to identify the daily stay-points. We achieve this task by performing extrema extraction on this curve.
 
\textbf{Extrema extraction}. Let's see what a stay-point looks like on a spatial curve. We remind that for every point $(x_i,y_i)$ of the spatial curve, $x_i$ is the minute, elapsed from midnight, that the cumulative distance $y_i$ is recorded. It is also necessary to remind that the spatial curve is an increasing curve. We claim that \textit{\textbf{the stay-points are the local minima of the $1^{st}$ derivative of the spatial curve}}; the rationale follows.

The first derivative of the spatial curve $C'(x,y)=\Delta y/\Delta x$ provides the speed of movement. So, when the user stops the speed is zero or  ideally $C'(x,y)=0$. An observation we make is that the $1^{st}$ derivative values are always greater than or equal to zero (fig.~\ref{fig:spatialCurveWithDerivatives}b) because the spatial curve is an increasing one (fig.~\ref{fig:spatialCurveWithDerivatives}a). Consequently it is easy to see that all stay-points are local minima of the $1^{st}$ derivative curve. These local minima can be identified using the $2^{nd}$ derivative (fig.~\ref{fig:spatialCurveWithDerivatives}c) check, which is $C''(x,y)=\Delta^2 y/\Delta x^2=0$. However, in practice the second derivative would never be zero. Instead, a numerically stable widely used approach is to identify these local minima by checking for zero-crossings of the $2^{nd}$ derivative~\cite{Haralick84}. The  $2^{nd}$ derivative will cross zero, from positive to negative, indicating deceleration until the stop. When the user starts moving again, the  $2^{nd}$ derivative will cross zero again, indicating acceleration due to the movement.

A drawback of the $2^{nd}$ derivative zero-crossing check is that even though this check is adequate for identifying extrema of the spatial curve, it also identifies inflection points of the curve. \textit{Inflection points} are stationary points where the sign of curve's curvature changes (or the curve's concavity changes). Given what the spatial curve represents, inflection points may indicate short stops, e.g. due to traffic jams.

Generally it is easy to decide between an extremum or an inflection point by just checking the concavity of the curve or by thresholding on the $1^{st}$ derivative value because the $1^{st}$ derivative is in the vicinity of zero ($C'(x,y)\approx 0$) only for extrema points and conversely it is far larger than zero for inflection points. However, in this application where positioning errors are abundant and the spatial curve is not smooth, these approaches cannot be applied without some error filtering. As smoothing the curve is costly and using hard thresholding iterates the problem of having to identify a good threshold, we decide to use soft thresholding by using a confidence value, as explained later.

\textbf{Stay-point region.} It is evident that the zero-crossings of the $2^{nd}$ derivative return a pair of points ($\mathbf{p}_i,\mathbf{p}_j)$, where $i<=j$. Even though we may get a pair of points we could also get a single point, if $i=j$. In addition, we could get more points (when $i-1<j$), i.e. a stay-point region $[i, j]$. This is completely normal as it depends on the number of locations returned by the trajectory tracking technique when the user is not moving. Considering the scenario that the user has slowed down significantly, let's say due to traffic, closeby the destination, before completely stopping, it is possible that this region contains outlier points. Similarly, it is possible that the whole region corresponds to inflection points.

\textbf{Extremum or Inflection?} Given a pair of points ($\mathbf{p}_i,\mathbf{p}_j)$ that corresponds to consecutive zero-crossings, the aim is to compute a confidence value that provides insight into whether the corresponding region $[i,j]$ is an extremum region.

First we compute the confidence value $C(\mathbf{p}_w)$ for each point $\mathbf{p}_w$, $i \le w \le j$ according to equation \ref{eq:confidence}, where $\mathbf{p}^{'}_{w}$ is the $1^{st}$ derivative on point $\mathbf{p}_w$, $\mathbf{p}^{'}_{min}$ is the minimum $1^{st}$ derivative over all points of the curve. Equation~\ref{eq:confidence} sets the confidence to $100\%$ when the first derivative is less than $e$. When the first derivative's value is increasing the confidence decreases until it becomes zero. Our experiments showed that a good value for $e$ is $0.05$ as this doesn't produce false minima. However, $e$ could be even smaller. A smaller $e$ can be used to reduce error due to noise, if such exists, by making it stricter to identify a stay-point.

The second step is to compute the confidence $C(\mathbf{p}_i,\mathbf{p}_j)$ for the whole region (see eq.~\ref{eq:totalConfidence}). The confidence value for the region $[i, j]$ is the confidence of the subregion $[k, w]$ (see eq.~\ref{eq:partialConfidence}) with maximum confidence value.

\begin{equation}
\begin{split}
C(\mathbf{p}_w) &=100\mbox{, if }\mathbf{p}_w^{'}<=e\\
                           &=100-\frac{\mathbf{p}^{'}_w-e}{\mathbf{p}^{'}_{min}}\mbox{, if }\mathbf{p}^{'}_w>e\\
                           &=0\mbox{, if }\frac{\mathbf{p}^{'}_w-e}{\mathbf{p}^{'}_{min}}>100
\end{split}
\label{eq:confidence}
\end{equation}

\begin{equation}
C(\mathbf{p}_k,\mathbf{p}_w)=\frac{\sum_{i=k}^w C(\mathbf{p}_i)}{w-k+1}
\label{eq:partialConfidence}
\end{equation}

\begin{equation}
C(\mathbf{p}_i,\mathbf{p}_j)=\max_{i\le k\le w\le j} C(\mathbf{p}_k,\mathbf{p}_w)
\label{eq:totalConfidence}
\end{equation}

After the computation of the confidence $C(\mathbf{p}_i,\mathbf{p}_j)$ we classify the region $[k, w]$ as a stay-point region when $C(\mathbf{p}_i,\mathbf{p}_j)\ge 80$. When the confidence falls in the range $[60,80)$ the region is considered a candidate stay-point region and we discard regions with lower confidence values. A candidate stay-point region may become a stay-point if the user identifies it as such or if an analysis that aims to identify user stay-point patterns indicates that it is a stay-point. Such an analysis is out of the scope of the current work.

\subsection{Computational Complexity} 
The proposed method requires the computation of the $1^{st}$ and $2^{nd}$ derivatives and the computation of a maximum confidence value for each candidate region. We compute the derivatives using central differences with $2^{nd}$ order accuracy. Hence the derivatives can be computed concurrently. The confidence value requires $p^{'}_{min}$, which is the minimum $1^{st}$ derivative over all points of the curve. Similarly this can be obtained during the computation of the $1^{st}$ derivative. In addition, the maximum confidence value for a region can be computed in linear time as computationally this would mean to compute the confidence value for the region and then trim from left and right all points such that confidence increases. Therefore the computational complexity of the proposed method is $O(N)$, where $N$ is the number of points of the spatial curve. 

In addition, replacing $\mathbf{p}^{'}_{min}$ with the constant $0.001$, which numerically is a good one, relaxes the need to have all $1^{st}$ derivatives prior to computing $\mathbf{p}^{'}_{min}$. The advantage is that the proposed method becomes real-time as it is able to process trajectory points as they arrive and in this way identify stay-points as they appear.   

\section{Experimental Evaluation}
\label{sec:evaluation}
The objective of the evaluation is to examine if the proposed work performs in line to the theoretical analysis and identify factors that may affect performance.

To this end, we have collected daily spatial trajectories for a period of $28$ days. This data was collected using different spatial tracking techniques. We executed the proposed method on the data produced by these techniques and collected all automatically generated stay-points with their confidence values. In order to make a proper evaluation, we compared them with the ground truth, i.e. the actual stay-points. 

It is important to mention that we don't perform an evaluation on publicly available datasets such as GeoLife~\cite{Zheng10} because the ground truth is not provided. As a result it is not possible to properly evaluate the performance of our method.

\subsection{Data Acquisition}
\label{subsec:dataAcquisition}
The iOS operating system via the coreLocation framework provides two services for spatial tracking: (a) the standard location service and (b) the significant location service. Service (a) relies on the GPS sensor, generates uniformly spaced data with various densities. Service (b) relies on the cell towers, generates non-uniformly spaced, sparse data, as it provides a new location at least every $500m$ or $5$ minutes. 

\begin{figure}[h]
  \centering\begin{tabular}{c}
     \begin{minipage}{8cm}
     \centering
     \includegraphics[width=7cm]{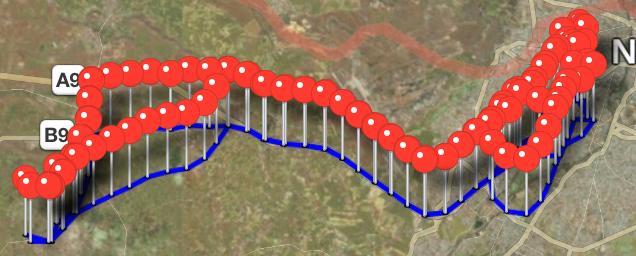}
    \end{minipage}
\\
     \begin{minipage}{8cm}
     \centering
     \includegraphics[width=7cm]{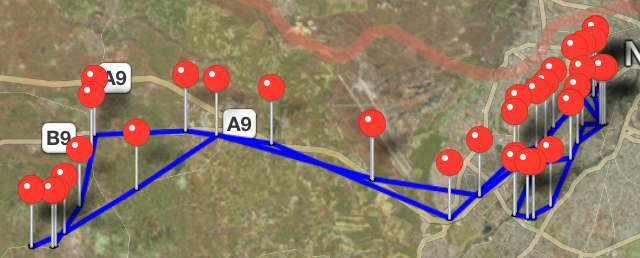}
   \end{minipage} 
\end{tabular}
\caption{Trajectory tracking examples for a single day. Top: SLS-500. Bottom: Hybrid.}
\label{fig:example}
\end{figure}

We implemented three applications that use the standard location service in order to acquire the trajectory path using three different densities: $100$, $250$ and $500$ meters. We denote these as $SLS-100$, $SLS-250$ and $SLS-500$. In addition, we implemented one application that uses a combined approach~\cite{Stylianou15}, which is denoted as \textit{hybrid}. This approach uses primarily the significant location service for spatial tracking. Whenever a new location is acquired it schedules the standard location service to power up in one minute in order to capture extra locations; this ensures that if a stay-point occurred the correct position of this point is captured. Examples of trajectories generated by two different techniques for the same day are provided in figure~\ref{fig:example}. We quickly observe the differences in data density and uniformity. 

The trajectory points were acquired using these four different iOS applications running on an iphone 6S and an iphone 5. Both devices run the iOS $10.x$ operating system. The iPhone 6S was used to run the $SLS-500$ and the hybrid and the iPhone 5 was used to run the $SLS-100$ and $SLS-250$. In total, we acquired trajectories for $60$ days and $28$ unique days. We acquired trajectories for $8$ days using all four techniques, $8$ days using two techniques ($SLS-500$, hybrid) and $12$ days using only the hybrid approach.

It is important to mention that the acquired data do not go through any noise removal process for the purposes of increasing accuracy rates. This means that the trajectory data exhibit noise from $10m$, which is normal, to about $1000$ meters, which is extreme. The underlying reason is that we want to demonstrate, via the experiments, robustness to noise.

\subsection{Ground Truth}
\label{subsection:groundtruth}
\begin{figure}[h]
  \centering\begin{tabular}{c}
     \begin{minipage}{7cm}
     \centering
     \includegraphics[width=7cm]{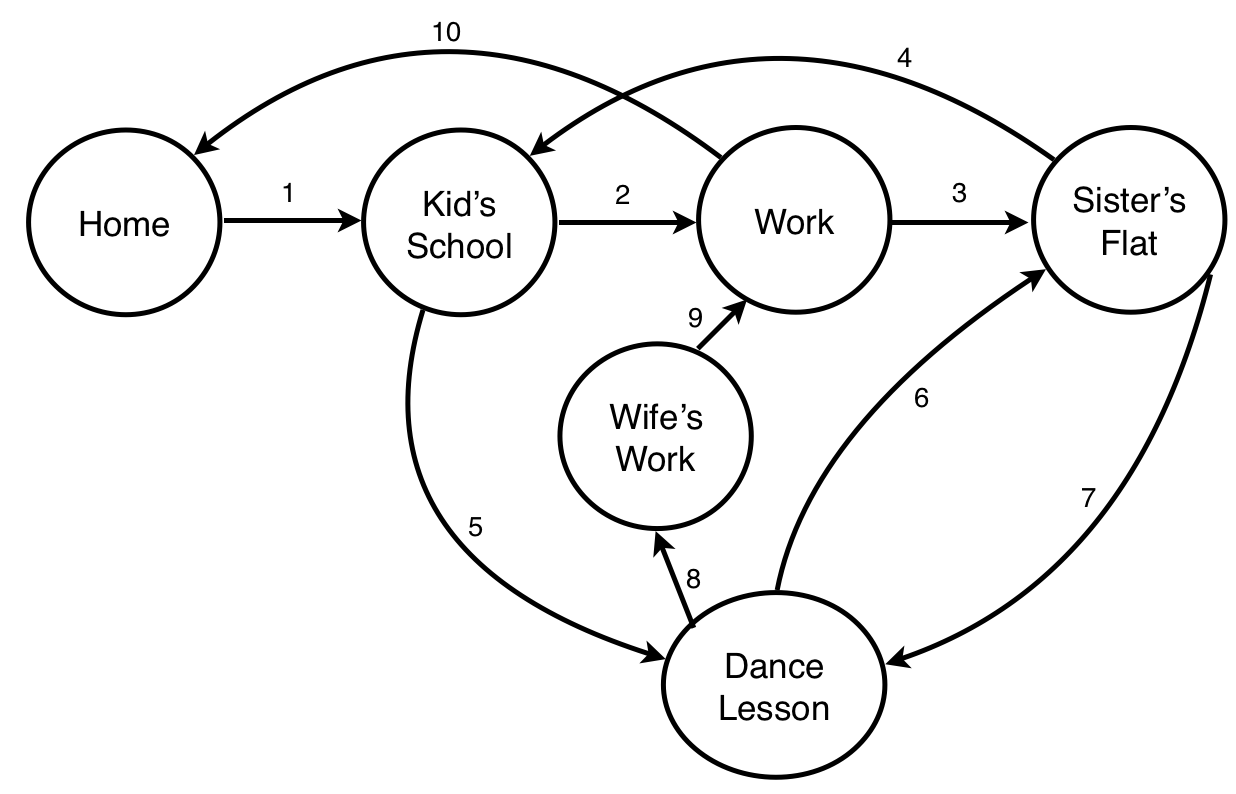}
   \end{minipage} 
\end{tabular}
\caption{One typical Wednesday.}
\label{fig:typical_day}
\end{figure}

During the evaluation the subject manually logged in the end of each day its stay-points including approximate arrival and departure times.  Figure~\ref{fig:typical_day} shows a state diagram for the subject's stay-points for a single day. The subject was moving in urban and suburban areas by car and was concurrently using the two iPhones to automatically track its daily whereabouts.

The ground truth stay-points are utilized in two ways: (a) we manually check whether these were captured in the raw trajectory data and (b) we compare them to the automatically identified stay-points in order to evaluate the accuracy of the proposed method.

\subsection{Results}
\label{subsection:comparison}
In order to evaluate the accuracy of the proposed method, we compared the stay-points extracted from the raw trajectory data, acquired using four different techniques, to the ground truth stay-points. The comparison aims to reveal the accuracy for automatic stay-point identification, the accuracy for inflection point identification, the percentage of false positive stay-points. In addition, we compared the automatically approximated stay-point duration to the actual duration and also examined whether a time threshold for identifying stay-points could exhibit similar accuracy to the proposed method.

\begin{table}[htb]
\centering
\caption{Summary of results. It includes the number of trajectory days, average number of trajectory points per day, the success rate and the percentage of false positives.}
\begin{tabular}{|l | p{1cm}| p{1.5cm}| p{1.5cm}| p{1.5cm}|}
\hline
         & \# of stay-points & Points per day & Success rate (\%) & False positive (\%) \\ \hline
Hybrid   & 180         & 48             & 86                & 2.8                 \\ \hline
SLS-100 & 56        & 97             & 98                & 13                  \\ \hline
SLS-250 & 56        & 78             & 84                & 9                   \\ \hline
SLS-500 & 108         & 54             & 60                & 5                   \\ \hline
\end{tabular}
\label{table:summaryResults}
\end{table}

Table~\ref{table:summaryResults} shows the overall success rate of stay-point identification for all data acquisition techniques. The best accuracy is achieved by the $SLS-100$ technique with about $98\%$ success rate, and the second best accuracy by the hybrid approach with $86\%$ success rate.  In addition, we observe that our proposed method exhibits a very few number of false positives -these are falsely identified stay-points-, which can be as low as $2.8\%$. 
Since the hybrid method produces the least number of data points, we believe that the reduced data density reduces the number of false positives and that the stay-point identification accuracy is not directly related to the density of the data points.

We also observe that given adequate data points the results could reach up to nearly $100\%$ accuracy rates. However, we notice that the increased number of data points increases the possibility for false positives as these increase up to $13\%$. The results show that the false positives and the missed stay-points are not repeating over the same days (e.g. all Tuesdays). Therefore, a good method that infers the user's repeating patterns will not include most of false positives and will increase the success rate for stay-points due to inference of the missing ones. After studying the results, we conclude that false positives occurred mainly due to urban traffic and very rarely due to other reasons. The increasing number of data points produced by the techniques $SLS-500$, $SLS-250$ and $SLS-100$, captured urban traffic with increasing accuracy. In one such case, the author's (driver) car was stationary for more $15$ minutes. Unfortunately, such cases are quite difficult to identify automatically without incorporating a learning method into the process.

It is as important to investigate the reasons for failing to identify some stay-points. According to table~\ref{table:missingStaypoints}, several stay-points were not identified by the hybrid method because (a) there were no trajectory points captured in their vicinity ($3.3\%$), (b) there were very few trajectory points in the vicinity of stay-points ($5.6\%$), (c) the stay was very short ($2.2\%$) of about a minute or so or (d) stay-points were identified but with lower than $80\%$ confidence ($1.7\%$). If we assume non-existence of reasons due to trajectory tracking techniques then we can deduce that the success rate of our proposed method is at least $90\%$, for the hybrid method. This hypothesis could theoretically increase the success rate for $SLS-250$ and $SLS-500$ to about $85\%$. Given the fact that the proposed method executed on raw trajectory data, we believe that the identification rate is quite satisfactory and can increase due to pre-processing. Pre-processing can help to reduce the effect of erroneous data as the trajectory data, in some cases, exhibited positioning errors from $5m$, which is normal, to more than $1000m$, which is abnormal, both in urban and suburban areas.

In addition, the results summarized in both tables~\ref{table:summaryResults} and~\ref{table:missingStaypoints} show that the proposed formula for computing the confidence value has very good results. First, only very few stay-points are not identified due to low confidence values; most of these are candidate stay-points. Second, the number of false positives due to high confidence can be as low as $2.8\%$. We note that the main reason for increased false positives e.g. in $SLS-100$ was not the confidence value, as explained before. Third, the proposed method assigned correct low confidence value to a total of $204$ inflection points, acquired by all four techniques, which are $50\%$ of the total number of stay-points. 

\begin{table}[htb]
\centering
\caption{The reasons some stay-points were not identified.}
\label{table:missingStaypoints}
\begin{tabular}{|l | p{1cm}| p{1.5cm}| p{1.5cm}| p{1.5cm}|}
\hline
                               & Short stay (\%)           & Few points (\%)           & No points (\%)            & Low confidence (\%)       \\ \hline
Hybrid  & 2.2 & 5.6&3.3 & 1.7 \\ \hline
SLS-250 & 1.8 & 0   & 7.3&1 \\ \hline
SLS-500                       & 0.9                      & 3.7                      & 21.3                     & 0                        \\ \hline
\end{tabular}
\end{table}

Further, we have evaluated the accuracy of the automatically computed duration by the two best performing trajectory tracking techniques for all the common stay-points and days by comparing them to the actual duration. Given the stay-point region $[k,w]$, the duration of stay was estimated by considering the time to distance from $\mathbf{p}_{k-1}$ to $\mathbf{p}_k$ and from $\mathbf{p}_{w}$ to $\mathbf{p}_{w+1}$ in addition to the time between $\mathbf{p}_k$, $\mathbf{p}_w$. The time to distance was estimated assuming traveling speed of $50Km/h$. The actual duration ranges from $5$ minutes to several hours. Table~\ref{table:durationComparison} summarizes the comparison. Both trajectory tracking techniques over-estimated the duration by about $19\%$ and $13\%$ on average. These deviations are on average about $12.5$ minutes for both techniques. However, the standard deviation shows that the deviations from the average for the $SLS-100$ technique are smaller than the hybrid technique. This makes sense because $SLS-100$ produces twice the number of data points to the hybrid technique.  

\begin{table}[htb]
\centering
\caption{Comparisons of the estimated duration for 43 stay-points using trajectory data from two different techniques. The actual total duration is 4194 minutes.}
\label{table:durationComparison}
\begin{tabular}{|l|p{1.5cm}|p{1cm}|p{1cm}|p{1cm}|p{1cm}|}
\hline
 &       & \multicolumn{4}{c|}{Duration (deviation from actual)}                       \\ \cline{2-6} 
                  & Total Minutes & Average (\%) & STD (\%) & Average (min) & STD (min) \\ \hline
Hybrid            & 4517  & 19           & 30        & 12.6          & 20.2      \\ \hline
SLS-100           & 4294  & 13           & 17        & 12.5          & 16.7      \\ \hline
\end{tabular}
\end{table}

Finally given the estimated duration, we examined whether a simple threshold could be adequate as most of the previous work suggests. For instance, one could claim that since, according to our experiments, the minimum actual duration is $5$ minutes then it would be adequate to just use a time threshold of $5$ minutes in order to identify a stay-point. To this end, we have again examined the duration of all the inflection points for the best performing techniques. We remind that an inflection point is an extremum point (or region) that is assigned a confidence value less than $60\%$. 

Our findings, which are summarized in table~\ref{table:inflectionPointDurations}, reveal that the minimum, maximum and average duration for the hybrid technique are $4$,  $49$ and $9$ minutes, respectively. In addition, the minimum, maximum and average duration for the $SLS-100$ technique are $2$, $24$ and $4$ minutes, respectively. The aforementioned results demonstrate that there is a significant gap between the minimum and maximum duration. Probably the most interesting outcome is that a time threshold of $5$ minutes~\cite{Nishida14} would label $30$ and $16$ inflection points as stay-points for the hybrid and $SLS-100$ techniques, respectively. Consequently, the accuracy rates would have been significantly smaller as false positives will increase $6$ times and $2.5$ times for the hybrid and $SLS-100$ techniques, respectively.

\begin{table}[htb]
\centering
\caption{The duration of inflection points for the Hybrid and SLS-100.}
\label{table:inflectionPointDurations}
\begin{tabular}{|l|l|l|l|l|l|}
\hline
 & \#of Points & \multicolumn{4}{c|}{Duration} \\ \cline{3-6} 
                  &                              & Min     & Max    & Average   & $\ge 5'$ \\ \hline                  
Hybrid            & 32                           & 4       & 49     & 9   &   30  \\ \hline
SLS-100           & 54                           & 2       & 24     & 4   &  16   \\ \hline
\end{tabular}
\end{table}

\section{Conclusion and Future Work}
We have presented a novel method that identifies stay-points using a geometry-based approach. In a nutshell, our method transforms the initial trajectory path to a time series curve such that the stay-point identification problem becomes an extrema extraction problem. The proposed method avoids using experimentally deduced distance, time or other threshold values. In contrast, it uses a confidence formula in order to robustly distinguish stay-points from stationary points. It is also runs in linear time and can run in real time.

We evaluated our method on raw trajectory data, for $28$ unique days, acquired using four different techniques with varying densities. 
The results demonstrate that the proposed method has $85-98\%$ accuracy and it is insensitive to the trajectory point uniformity, density and noise. The confidence value, computed for each stay-point region, correctly identifies all stationary points and generates very few false positive stay-points. 

In the future, we aim to further improve the proposed method's accuracy by investigating various ways for either improving the quality of the raw trajectory data or by improving current trajectory tracking techniques. A way to do either is to incorporate a machine learning method. In addition, we plan to extend our method to $3D$ such that we can identify user patterns using $3D$ feature extraction.

\bibliographystyle{plain}
\bibliography{bibpaper}  


\begin{thebibliography}{00}


\ifx \showCODEN    \undefined \def \showCODEN     #1{\unskip}     \fi
\ifx \showDOI      \undefined \def \showDOI       #1{{\tt DOI:}\penalty0{#1}\ }
  \fi
\ifx \showISBNx    \undefined \def \showISBNx     #1{\unskip}     \fi
\ifx \showISBNxiii \undefined \def \showISBNxiii  #1{\unskip}     \fi
\ifx \showISSN     \undefined \def \showISSN      #1{\unskip}     \fi
\ifx \showLCCN     \undefined \def \showLCCN      #1{\unskip}     \fi
\ifx \shownote     \undefined \def \shownote      #1{#1}          \fi
\ifx \showarticletitle \undefined \def \showarticletitle #1{#1}   \fi
\ifx \showURL      \undefined \def \showURL       #1{#1}          \fi

\bibitem{Ashbrook03}
{Daniel Ashbrook} {and} {Thad Starner}. 2003.
\newblock \showarticletitle{Using GPS to Learn Significant Locations and
  Predict Movement Across Multiple Users}.
\newblock {\em Personal and Ubiquitous Computing\/}  {7(5)} (2003), 275--286.
\newblock


\bibitem{Chon14}
{Yohan Chon}, {Elmurod Talipov}, {Hyojeong Shin}, {and} {Hojung Cha}. 2014.
\newblock \showarticletitle{SmartDC: Mobility Prediction-based Adaptive Duty
  Cycling for Everyday Location Monitoring}.
\newblock {\em IEEE Transactions on Mobile Computing\/} 3 (2014), 512--525.
\newblock


\bibitem{Haralick84}
{Robert~M Haralick}. 1984.
\newblock \showarticletitle{Digital Step Edges from Zero Crossing of Second
  Directional Derivatives}.
\newblock {\em IEEE Transactions on Pattern Analysis and Machine
  Intelligence\/} 1 (1984), 58--68.
\newblock


\bibitem{Hariharan04}
{Ramaswamy Hariharan} {and} {Kentaro Toyama}. 2004.
\newblock \showarticletitle{Project Lachesis: Parsing and Modeling Location
  Histories}.
\newblock In {\em Geographic Information Science}. Springer, 106--124.
\newblock


\bibitem{Herder14}
{Eelco Herder}, {Patrick Siehndel}, {and} {Ricardo Kawase}. 2014.
\newblock \showarticletitle{Predicting User Locations and Trajectories}.
\newblock In {\em User Modeling, Adaptation, and Personalization}. Springer,
  86--97.
\newblock


\bibitem{Horanont14}
{Teerayut Horanont}, {Santi Phithakkitnukoon}, {and} {Ryosuke Shibasaki}. 2014.
\newblock \showarticletitle{Sensing urban Density using Mobile Phone GPS
  Locations: A Case Study of Odaiba Area, Japan}.
\newblock In {\em Nature of Computation and Communication}. Springer, 146--155.
\newblock


\bibitem{Kang05}
{Jong~Hee Kang}, {William Welbourne}, {Benjamin Stewart}, {and} {Gaetano
  Borriello}. 2005.
\newblock \showarticletitle{Extracting Places from Traces of Locations}.
\newblock {\em ACM SIGMOBILE Mobile Computing and Communications Review\/} {9},
  3 (2005), 58--68.
\newblock


\bibitem{Kjaergaard11}
{Mikkel~Baun Kj{\ae}rgaard}, {Sourav Bhattacharya}, {Henrik Blunck}, {and}
  {Petteri Nurmi}. 2011.
\newblock \showarticletitle{Energy-Efficient Trajectory Tracking for Mobile
  Devices}. In {\em 9th Int. Conf. on Mobile Systems, Applications, and
  Services}. ACM, 307--320.
\newblock


\bibitem{Nishida14}
{Kyosuke Nishida}, {Hiroyuki Toda}, {Takeshi Kurashima}, {and} {Yoshihiko
  Suhara}. 2014.
\newblock \showarticletitle{Probabilistic Identification of Visited
  Point-of-Interest for Personalized Automatic Check-in}. In {\em ACM Int.
  Joint Conf. on Pervasive and Ubiquitous Computing}. 631--642.
\newblock


\bibitem{Palma08}
{Andrey~Tietbohl Palma}, {Vania Bogorny}, {Bart Kuijpers}, {and} {Luis~Otavio
  Alvares}. 2008.
\newblock \showarticletitle{A Clustering-based Approach for Discovering
  Interesting Places in Trajectories}. In {\em ACM symposium on Applied
  computing}. 863--868.
\newblock


\bibitem{Pavan15}
{Marco Pavan}, {Stefano Mizzaro}, {Ivan Scagnetto}, {and} {Andrea Beggiato}.
  2015.
\newblock \showarticletitle{Finding Important Locations: A Feature-Based
  Approach}. In {\em 16th IEEE Int. Conf. on Mobile Data Management}, Vol.~1.
  110--115.
\newblock


\bibitem{Ranvier15}
{Jean-Eudes Ranvier}, {Michele Catasta}, {Matteo Vasirani}, {and} {Karl
  Aberer}. 2015.
\newblock \showarticletitle{RoutineSense: A Mobile Sensing Framework for the
  Reconstruction of User Routines}. In {\em 12th Int. Conf. Mobile and
  Ubiquitous Systems: Computing, Networking and Services}.
\newblock


\bibitem{Stylianou15}
{Georgios Stylianou}. 2015.
\newblock \showarticletitle{Efficient Spatial Tracking for iOS Devices}. In
  {\em 12th EAI Int. Conf. on Mobile and Ubiquitous Systems: Computing,
  Networking and Services}.
\newblock


\bibitem{Thomason15}
{Alasdair Thomason}, {Nathan Griffiths}, {and} {Victor Sanchez}. 2015.
\newblock \showarticletitle{Identifying Locations from Geospatial
  Trajectories}.
\newblock {\it J. Comput. System Sci.} (2015).
\newblock


\bibitem{Yadav14}
{Kuldeep Yadav}, {Vinayak Naik}, {Abhishek Kumar}, {and} {Prateek Jassal}.
  2014.
\newblock \showarticletitle{PlaceMap: Discovering Human Places of Interest
  Using Low-Energy Location Interfaces on Mobile Phones}. In {\em 5th ACM
  Symposium on Computing for Development}. 93--102.
\newblock


\bibitem{Ye09}
{Yang Ye}, {Yu Zheng}, {Yukun Chen}, {Jianhua Feng}, {and} {Xing Xie}. 2009.
\newblock \showarticletitle{Mining Individual Life Pattern Based on Location
  History}. In {\em 10th Int. Conf. on Mobile Data Management: Systems,
  Services and Middleware}. IEEE, 1--10.
\newblock


\bibitem{Zheng10a}
{Vincent~W Zheng}, {Yu Zheng}, {Xing Xie}, {and} {Qiang Yang}. 2010.
\newblock \showarticletitle{Collaborative Location and Activity Recommendations
  with GPS History Data}. In {\em 19th Int. Conf. on World Wide Web}. ACM,
  1029--1038.
\newblock


\bibitem{Zheng10}
{Yu Zheng}, {Xing Xie}, {and} {Wei-Ying Ma}. 2010.
\newblock \showarticletitle{GeoLife: A Collaborative Social Networking Service
  among User, Location and Trajectory}.
\newblock {\em IEEE Data Engineering Bulletin\/}  {33(2)} (2010), 32--39.
\newblock


\bibitem{Zheng09}
{Yu Zheng}, {Lizhu Zhang}, {Xing Xie}, {and} {Wei-Ying Ma}. 2009.
\newblock \showarticletitle{Mining interesting Locations and Travel Sequences
  from GPS Trajectories}. In {\em 18th Int. Conf. on World Wide Web}. ACM,
  791--800.
\newblock


\bibitem{Zhou14}
{Jun Zhou}, {Qinpei Zhao}, {and} {Hongyu Li}. 2014.
\newblock \showarticletitle{Integrating Time Stamps into Discovering the Places
  of Interest}.
\newblock In {\em Intelligent Computing Methodologies}. Springer, 571--580.
\newblock


\end{thebibliography}

\end{document}